\documentclass[namedreferences]{solarphysics}

\usepackage[hyperref,optionalrh]{spr-sola-addons} 
\usepackage{graphicx}        
\usepackage{color}           
\usepackage{breakurl}        
\usepackage{enumerate} 
\usepackage{comment}

\newcommand{\etal}{{\it et al.}}

\renewcommand{\vec}[1]{{\mathbfit #1}}
\newcommand{\deriv}[2]{\frac{{\mathrm d} #1}{{\mathrm d} #2}}

\newcommand{\aap}{    {\it Astron. Astrophys.}}

\newcommand{\apj}{    {\it Astrophys. J.}}

\newcommand{\cjaa}{   {\it Chin. J. Astron. Astrophys.}}

\newcommand{\mnras}{  {\it Mon. Not. Roy. Astron. Soc.}}

\newcommand{\pasj}{   {\it Pub. Astron. Soc. Japan}}

\newcommand{\solphys}{{\it Solar Phys.}}
 
\newcommand{\ssr}{    {\it Space Sci. Rev.}} 
\chardef\us=`\_

\begin{document}

\begin{article}
\begin{opening}

\title{Improvements of the Longitudinal Magnetic Field Measurement from the \textit{Solar Magnetic Field Telescope} at the Huairou Solar Observing Station}

\author[addressref={aff1,aff2},corref,email={plotnikov.andrey.alex@yandex.ru}]{\inits{A.A.}\fnm{Andrei}~\lnm{Plotnikov}}
\author[addressref={aff1,aff2}]{\inits{A.S.}\fnm{Alexander}~\lnm{Kutsenko}\orcid{0000-0002-1196-5049}}
\author[addressref={aff1,aff3}]{\inits{S.B.}\fnm{Shangbin}~\lnm{Yang}}
\author[addressref={aff1}]{\inits{H.Q.}\fnm{Haiquing}~\lnm{Xu}}
\author[addressref={aff1}]{\inits{X.Y.}\fnm{Xianyong}~\lnm{Bai}\orcid{0000-0003-2686-9153}}
\author[addressref={aff1}]{\inits{H.Q.}\fnm{Hongqi}~\lnm{Zhang}}
\author[addressref={aff4}]{\inits{K.M.}\fnm{Kirill}~\lnm{Kuzanyan}\orcid{0000-0003-1677-4417}}

\address[id=aff1]{Key Laboratory of Solar Activity, National Astronomical Observatories of China (Chinese Academy of Sciences), Beijing, 100101, China}
\address[id=aff2]{Crimean Astrophysical Observatory (Russian Academy of Sciences), Nauchny, Crimea, 298409, Russia}
\address[id=aff3]{University of Chinese Academy of Sciences, 100049, Beijing, China}
\address[id=aff4]{IZMIRAN, Russian Academy of Sciences, Troitsk, Moscow, 108840, Russia}

\runningauthor{A.A. Plotnikov \textit{et al.}}
\runningtitle{Improvements of the Longitudinal Magnetic Field Measurement at the HSOS/SMFT}

\begin{abstract}
The weak-field approximation, implying a linear relationship between Stokes-$V/I$ and longitudinal magnetic field [$B_{\Vert}$], often suffers from saturation observed in strong magnetic-field regions such as sunspot umbrae. In this work, we seek to improve the magnetic-field observations carried out by the \textit{Solar Magnetic Field Telescope} (SMFT) at the Huairou Solar Observing Station, China. We propose using a non-linear relationship between Stokes-$V/I$ and $B_{\Vert}$ to derive the magnetic field. To determine the form of the relationship, we perform a cross-calibration of the observed SMFT data and magnetograms provided by the \textit{Helioseismic and Magnetic Imager} onboard the \textit{Solar Dynamics Observatory}. The algorithm for the magnetic-field derivation is described in detail. We show that using a non-linear relationship between Stokes-$V/I$ and $B_{\Vert}$ allows us to eliminate magnetic field saturation inside the sunspot umbra. The proposed technique enables one to enhance the reliability of the SMFT magnetic-field data obtained even long before the space-based instrumentation era, since 1987. 
   
\end{abstract}
\end{opening}

\section{Introduction}
     \label{S-Introduction} 

The \textit{Solar Magnetic Field Telescope} (SMFT) is a narrow-band filtergraph installed at the Huairou Solar Observing Station (HSOS), the National Astronomical Observatories, Chinese Academy of Sciences \citep{Ai1987}. The field-of-view of the telescope is 225$^{\prime\prime}\times$170$^{\prime\prime}$. The pixel size of the acquired filtergrams is 0.295$^{\prime\prime}\times$0.287$^{\prime\prime}$ while the spatial resolution of about 2$^{\prime\prime}$ is limited by atmospheric seeing. The telescope operation of the SMFT provides a continuous homogeneous observational data set on the solar vector magnetic field since 1987. Such a long unique data series allows one to analyze long-term variations and solar cycle dependence of active region parameters that rely on the vector magnetic field measurements, for example current helicity of active regions \citep[e.g.][]{Zhangetal2010}.
SMFT uses the photospheric Fe~{\sc i} 5324.19\ \AA\ spectral line to derive information on solar magnetic fields. Stokes-$(V/I)_{\rm SMFT}$ are routinely measured at a single-wavelength in the wing of the spectral line ($\approx$75\ m\AA\ from the line center). 

The magnetic field is traditionally obtained in the framework of the weak-field approximation, implying a linear relationship between the measured Stokes $(V/I)_{\rm SMFT}$ and longitudinal magnetic field [$B_{\Vert}$]:

   \begin{equation}  \label{weak-field}
		B_{\Vert} = C_{\Vert} \frac{V}{I}.
	\end{equation}
	
	
The calibration coefficient [$C_{\Vert}$] for HSOS/SMFT measurements was analytically obtained for the first time by \citet{Ai1982}. Several other methods for weak-field calibration have been proposed recently \citep[see][and references therein]{Bai2013}.

In fact, the linear relationship between the Stokes-$V/I$ and longitudinal magnetic field $B_{\Vert}$ holds as long as the magnetic field does not exceed a certain value that depends on many factors. As we will show below, this value is about 1000\ G for SMFT. For stronger magnetic fields, which are readily observed in the umbrae of sunspots, the Stokes-$V/I$ decreases with increasing [$B_{\Vert}$] \citep[e.g.][]{Sakurai1995} resulting in significant underestimation of the sunspot umbra magnetic field. It is worth noting that the deviation from the linear relationship between the Stokes-$V/I$ and longitudinal magnetic field $B_{\Vert}$ is also caused by different solar atmosphere conditions \citep[e.g.][]{Zhang2019}. Thus, quiet-Sun areas and umbral regions exhibit different magnetic sensitivity due to temperature differences. The ratio between the Stokes-$V/I$ measured for quiet-Sun and within umbra regions may reach 1.8 for the Fe~{\sc i} 5324.19\ \AA\ spectral line \citep{Zhang2019}.

The reliability of the transverse magneti-field measurements also depends to a great extent on the measurements of longitudinal magnetic field. The reason for this is the so-called 180-degree ambiguity problem when the azimuthal direction of the transverse magnetic-field vector cannot be directly determined from the observations. A number of techniques have been proposed to solve this ambiguity \citep[e.g.][]{Metcalf1994}. Most of the techniques use the observed longitudinal magnetic field to calculate the configuration of the potential field. Then, the observed transverse magnetic field is directed along the potential transverse field \citep[e.g.][]{Canfield1993}. Other methods may use various criteria to perform the disambiguation. However, it is worth noting that the disambiguation procedure relies exclusively on a number of physical assumptions. Calculations of electric currents imply deriving the spatial derivative of the transverse magnetic field. In such a case, the magnitude, sign, and spatial structure of the electric currents and current helicity depend strongly on the correct 180-degree disambiguation of the transversal magnetic field. In the case of longitudinal magnetic-field saturation inside a sunspot, the 180-degree disambiguation procedure may  turn the azimuthal component of the transverse magnetic field in the wrong direction, resulting in incorrect estimation of electric currents and, consequently, of current helicity.

The saturation problem might be solved by applying state-of-the-art machine learning techniques. Thus, recently \citet{Guo2020} developed an approach for non-linear calibration of filter-based magnetographs. The approach employs a trained multilayer perceptron aimed at deriving magnetic-field parameters from Stokes-$I$, -$Q$, -$U$, and -$V$ measured at a single wavelength. However, the method is still to be adopted for HSOS/SMFT data.

The saturation of the longitudinal magnetic field inside a sunspot can be eliminated by using additional information on spectral line, i.e., when the data on spectral line profile are available. Thus, \cite{Bai2014} derived the magnetic-field vector by fitting six points of the Fe~{\sc i} 5324.19\ \AA\ spectral line profile by analytical Stokes profiles obtained in the Milne--Eddington (ME) atmosphere approximation. A similar procedure was applied by \cite{Su2004} who used 31 points of the same spectral line profile to derive the vector magnetic field. See Section~\ref{S-theory} for more details on the magnetic field deriving from the observed Stokes vector. The longitudinal magnetic field can also be derived by a more rapid and straightforward center-of-gravity method \citep{Rees1979}: $B_{\Vert}$ is proportional to the difference of the centers of gravity of spectral lines in right- and left-circular polarization states.

However, routine measurements of Stokes-$(V/I)_{\rm SMFT}$ by HSOS/SMFT are performed at a single wavelength point of Fe~{\sc i} 5324.19\ \AA\ spectral line profile. In this article, we make an attempt to improve the routine magnetic-field measurements by introducing a non-linear relationship between the Stokes-$(V/I)_{\rm SMFT}$ and longitudinal magnetic field.

\section{Theoretical Background}
\label{S-theory}

Polarization of the light emitted by the solar atmosphere emerges as a result of the radiation propagation through a magnetized plasma. The resultant state of the light is described by the wavelength-dependent Stokes pseudo-vector $\vec{I}=(I, Q, U, V)$, where $I$ is the total intensity, $Q$ and $U$ are components associated with linear polarization, and $V$ is the component attributed to circular polarization. The Stokes profiles determined by physical parameters of the atmosphere can be derived by solving the radiative transfer equation (RTE) for a certain atmosphere model. However, in practice the observables are four spectral profiles of the Stokes vector components $I(\lambda), Q(\lambda), U(\lambda), V(\lambda)$. Therefore, one has to solve the inverse problem to infer the magnetic field and thermodynamical parameters of the medium the light propagates through \citep[see, e.g.][]{delToroIniesta2016}. A description of currently used techniques and approaches used to inverse the Stokes profiles can be found in the review by \citet{delToroIniesta2016}. 

The RTE can be expressed in the form \citep[e.g.][]{delToroIniesta2003}
\begin{equation}
    \deriv{\vec{I}}{\tau_{c}} = \mathbf{K} (\vec{I} - \vec{S}),
\end{equation}
where $\tau_{c}$ is the optical depth at the continuum wavelength, $\mathbf{K}$ is the propagation matrix, and $\vec{S}$ is the source function. The propagation matrix $\mathbf{K}$ describes energy absorption, injection, and transfer between different polarization states \citep[e.g.][]{delToroIniesta2016}. Elements of the propagation matrix are combinations of Voight and Faraday--Voigt functions. In general, the propagation matrix $\mathbf{K}$ and the source function $\vec{S}$ depend on the optical depth.

The inversion of the RTE requires numerical solution of a set of integral equations. This problem requires a lot of resources when one considers a stratified atmosphere that is in a non-local-thermodynamic equilibrium state. However, for certain special cases an analytical solution exists. One of the most widely used simplifications is the ME atmosphere. ME assumes the propagation matrix to be independent of the optical depth and the source function to be linearly dependent on the optical depth, \textit{i.e.} ME is applicable to a certain extent for photospheric line analysis. In this case, the RTE has an analytical solution known as the Unno--Rachkovsky solution \citep{Unno1956, Rachkovsky1962}. The four Stokes profiles are determined by nine parameters, namely three components of the magnetic field (field strength, inclination, and azimuth), line-of-sight velocity of the plasma, Doppler width of the spectral line, line-to-continuum absorption coefficient, the damping parameter, the source function, and the source function's gradient. For instance, the ME approximation is used  to infer solar vector magnetic fields and Doppler velocities from Stokes-vector observations \citep{Borrero2011} taken by the \textit{Helioseismic and Magnetic Imager} onboard the \textit{Solar Dynamics Observatory} \citep[SDO/HMI: ][]{Scherrer2012, Schou2012}.

When the longitudinal magnetic field is weak and the Zeeman splitting of the line is much smaller than the Doppler width, the Voigt and Faraday--Voigt functions in the propagation matrix $\mathbf{K}$ can be replaced by the lower-order terms of its Taylor series \citep{LandiDeglInnocenti1973, Jefferies1989}. In such a case, assuming the atmosphere to be in local thermodynamic equilibrium, the RTE can be further simplified. \citet{LandiDeglInnocenti2004} showed that for weak magnetic-field regime the following relationship holds \citep[we use the notation from][]{delToroIniesta2016}:
\begin{equation}
    V(\lambda) = - g_{\rm eff} \Delta \lambda_{B} \cos\gamma \frac{\partial I_{\rm nm}}{\partial \lambda},
    \label{eq_wf}
\end{equation}
where $g_{\rm eff}$ is the effective Land{\'e} factor, $\gamma$ is inclination, $I_{\rm nm}$ is the Stokes $I$ profile in the absence of magnetic field, and the Zeeman splitting
\begin{equation}
    \Delta \lambda_{B} = \frac{\lambda^{2}_{0} \mathrm{e}_{0} B}{4 \pi \rm m c^{2}}.
    \label{eq_ZeemanSplitting}
\end{equation} 
In equation~\ref{eq_ZeemanSplitting}, $\lambda_{0}$ is the central wavelength of the spectral line, $\mathrm{c}$ is the speed of light, $\mathrm{e}_{0}$ and $\mathrm{m}$ are electron's charge and mass, respectively. By substituting equation~\ref{eq_ZeemanSplitting} into equation~\ref{eq_wf}, one can easily obtain the magnetograph formula (equation \ref{weak-field}). The weakness of the magnetic field implies
\begin{equation}
    g_{\rm eff} \frac{\Delta \lambda_{B}}{\Delta \lambda_{\rm D}} \ll 1,
    \label{wfa_validity}
\end{equation}
where $\Delta \lambda_{\rm D}$ is the Doppler width of the spectral line. However, the weak-field approximation is not valid for stronger magnetic fields: higher-order terms in the Taylor series of Voigt and Faraday--Voigt functions in the propagation matrix start to play a significant role as the magnetic-field strength increases. \citet{delToroIniesta2016} argued that the saturation of Stokes-$V/I$ can be seen even for relatively weak magnetic fields of about several hundreds of gauss.

Fe~{\sc i} 5324\ \AA\ spectral line used in HSOS/SMFT exhibits relatively large total width. One should expect the weak-field approximation to be valid within a broad range of magnetic field strength. The theoretical relationship between the Stokes-$V/I$ and longitudinal magnetic field for Fe~{\sc i} 5324\ \AA\ spectral line was investigated in \citet{Ai1982} and in \citet{Su2004}, and we refer the reader to those paper for details.

\begin{figure}[h]
	\centerline{\includegraphics[width=1.0\textwidth]{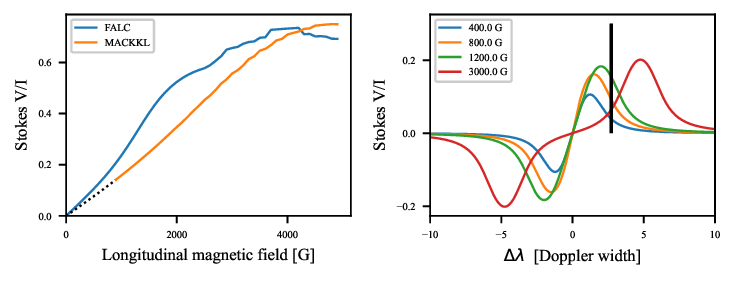}}
	\caption{
	Left -- Stokes-$V/I$ versus longitudinal magnetic-field relationship for Fe~{\sc i} 5324\ \AA\ spectral line. Stokes-$V/I$ is supposed to be sampled in the wing of the spectral line. The relationship was calculated using SIR code. The atmospheric thermodynamics parameters were retrieved from the FALC (blue curve) and MACKKL (orange curve) semi-empirical atmospheric models. The MACKKL model is applicable for sunspot umbrae, therefore the distribution for magnetic fields below 1000\ G is plotted by a dashed line. Right -- Simulated Stokes-$(V/I)$ profiles for several magnetic-field strength values. The calculations were performed in the framework of the ME approximation. The black vertical line shows the plausible tuning position of the filter used to sample Stokes-$(V/I)$. The reason of the saturation in the weak-field approximation is clearly seen. In both panels the magnetic field is co-aligned with the line-of-sight.
	}
	\label{Doppler_width_effect}
\end{figure}

In this work we also carried out the simulation of the saturation of the Stokes-$V/I$ in the weak-field regime for Fe~{\sc i} 5324\ \AA. The results are shown in the left panel of Figure~\ref{Doppler_width_effect}. The Stokes-$V$ and -$I$ profiles were calculated using SIR code \citep{RuizCobo1992, BellotRubio2003}. To roughly imitate filtergraph observations, the Stokes-vector profiles were sampled at a single wavelength position shifted by 75\ m\AA\ from the line center. The blue curve in the left panel of Figure~\ref{Doppler_width_effect} was derived by using atmospheric parameters provided by the semi-empirical FALC model \citep{Fontenla1993}. The model describes the quiet-Sun atmosphere; therefore, the left-hand (weak-field) part of the blue curve is supposed to be valid. The orange curve in the left panel of Figure~\ref{Doppler_width_effect} was obtained using atmospheric parameters from the MACKKL model by \citet{Maltby1986}. The latter model describes the thermodynamics of sunspot umbrae. Hence, this model could be more suitable for the strong-field part of the Stokes-$V/I$ versus $B_{\Vert}$ relationship. One can see that the curves have different slopes. In our opinion, simulation of the exact behaviour of Stokes-$V/I$ versus $B_{\Vert}$ in the weak-field regime requires more sophisticated atmospheric models taking into account changes of the atmosphere thermodynamic as one switches from weak quiet-Sun magnetic fields to strong magnetic field inside sunspot umbra. Besides that, the shape of the relationship also depends on other factors, e.g. on the magnetic field vector inclination \citep[see, e.g., Figure 11.1 of][]{LandiDeglInnocenti2004}.

The reason for the Stokes-$V/I$ saturation in the weak-field approximation can be easily seen in the right panel of Figure~\ref{Doppler_width_effect}. Colored curves show simulated Stokes-$V/I$ profile (in the framework of ME atmosphere approximation) for several values of the magnetic-field strength (the vector of the magnetic field is supposed to be co-aligned with the line-of-sight). The vertical black line denotes the central position of a filter in the wing of the spectral line. One can see that for weak magnetic fields the Stokes-$V/I$ sampled at the filter position increases gradually with magnetic field. As the magnetic field grows, the peaks of the $V/I$ profile shift farther from the spectral line center, resulting in saturation and further decreasing of sampled $V/I$. 

The saturation can be explained in a more simplified qualitative way. It is the Zeeman splitting of the spectral line, i.e. the displacement of the polarized component of the spectral line from the unperturbed-wavelength position, that is proportional to the magnetic-field strength (Equation~\ref{eq_ZeemanSplitting}). When the magnetic field is weak, the amplitude of the Stokes-$V/I$ is roughly proportional to the displacement. As the magnetic field strength increases, the amplitude of the Stokes-$V/I$ reaches some saturated value (the relationship $I^2 \ge Q^2 + U^2 + V^2$ holds) and shifts farther from the line center. Hence, in order to overcome the saturation, one has to use the information on the Stokes profile to determine the splitting itself. Therefore, inversion techniques such as those based on ME approximation or more sophisticated atmospheric models require several (more than two) wavelength points of the Stokes profiles as input data. At the same time, as stated by \citet{delToroIniesta2016}, the magnetograph formula (Equation \ref{weak-field}) is the only way to get information on the longitudinal magnetic field if the instrument samples the circular polarization at one or two wavelength positions.


	

\section{The Method}

To establish the relationship between the HSOS/SMFT Stokes-$(V/I)_{\rm SMFT}$ and longitudinal magnetic field [$B_{\rm SMFT}$] we performed a cross-calibration of Stokes-$(V/I)_{\rm SMFT}$  and longitudinal magnetic field $B_{\rm HMI}$ provided by SDO/HMI. SDO/HMI is a full-disk filtergraph that measures the profile of photospheric Fe~{\sc i} 6173\ \AA\ line at six wavelength positions in six polarization states to derive the information on four spectral profiles of the Stokes vector. The spatial resolution of the instrument is approximately 1.5$^{\prime\prime}$ with 0.5$^{\prime\prime}\times$0.5$^{\prime\prime}$ pixel size.


A ME-based Very Fast Inversion of the Stokes Vector code (VFISV) described in detail in \citet{Borrero2011} is used to process SDO/HMI pipeline data and to derive the vector magnetic field. The magnetic field maps of active regions are available in the form of patches of full-disk maps \citep{Bobra2014, Hoeksema2014}. These patches were used to perform cross-calibration with the HSOS/SMFT observations. SDO/HMI longitudinal magnetic field was derived as $B_{\rm HMI} = B^{\rm strength}_{\rm HMI} \cos(B^{\rm incl}_{\rm HMI})$, where  $B^{\rm strength}_{\rm HMI}$ and $B^{\rm incl}_{\rm HMI}$ are magnetic filed strength and inclination, respectively.

For cross-calibration we used co-temporal magnetograms of ten randomly selected active regions observed between 2015 and 2018. Magnetograms acquired by SDO/HMI were rotated by the $p$-angle and rescaled to the pixel size of HSOS/SMFT of approximately 0.29$^{\prime\prime}\times$0.29$^{\prime\prime}$ by a cross-correlation technique. Since SDO/HMI data are not affected by seeing, the magnetograms from the space borne instrument were smoothed by a 2D Gaussian kernel of 1.5$^{\prime\prime}\times$1.5$^{\prime\prime}$ to roughly imitate the atmospheric blurring. Then, the same regions of the solar surface were cropped from the processed SDO/HMI magnetograms and from HSOS/SMFT Stokes-$(V/I)_{\rm SMFT}$ maps.

\begin{figure}
	\centerline{\includegraphics{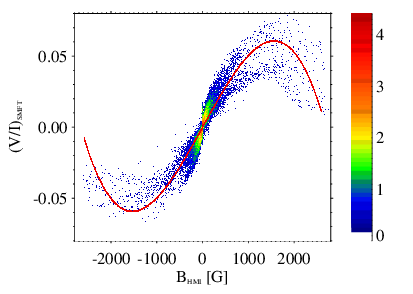}}
	\caption{The logarithmic density of HSOS/SMFT Stokes-$(V/I)_{\rm SMFT}$ versus SDO/HMI longitudinal magnetic field $B_{\rm HMI}$ distribution of ten randomly selected active regions observed between 2015 and 2018. The red calibration curve shows the best third order polynomial approximation (Equation~\ref{polynom}) of the distribution.
	}
	\label{fig1}
\end{figure}

The HSOS/SMFT Stokes-$(V/I)_{\rm SMFT}$ versus SDO/HMI $B_{\rm HMI}$ distribution for all the magnetograms used for the cross-calibration is shown in Figure~\ref{fig1}. The red calibration curve in the plot is the best least-square approximation of the distribution by a third-order polynomial
\begin{equation} \label{polynom}
    (V/I)_{\rm SMFT} = C_0 + C_1 \times B_{\rm HMI} + C_2 \times (B_{\rm HMI})^2 + C_3 \times (B_{\rm HMI})^3,
\end{equation}
where $C_0 = -1.030 \times 10^{-4}$, $C_1 = 5.815 \times 10^{-5}$ G$^{-1}$, $C_2 = 2.743 \times 10^{-10}$ G$^{-2}$, $C_3 = -8.061 \times 10^{-12}$ G$^{-3}$. One can see in Figure~\ref{fig1} that the quasi-linear relationship holds for relatively weak magnetic fields while Stokes-$(V/I)_{\rm SMFT}$ starts to saturate for longitudinal magnetic field exceeding approximately 1000~G. The calibration constant $C_{\Vert}$ in Equation~\ref{weak-field} can be easily derived as $C_{\Vert}=1/C_1$. We will refer to this constant as $C_{\rm SMFT}$ in the rest of the article. The value $C_1=5.815\times10^{-5}~ G^{-1}$ for the linear relationship between $(V/I)_{\rm SMFT}$ and $B_{\rm SMFT}$ is approximately 1.6 times smaller than the theoretical value obtained by \citet{Ai1982}. However, this difference can be attributed to the different spectral lines used by HSOS/SMFT and SDO/HMI.

Note that using the non-linear relationship between Stokes-$(V/I)$ and $B_{\Vert}$ has been proposed earlier. For example, \cite{Chae2007} performed cross-calibration of \textit{Narrow-band Filter Imager} (NFI) Stokes-$(V/I)$ and longitudinal magnetic field acquired by the \textit{Spectropolarimeter} (SP) onboard \textit{Hinode} \citep{Kosugi2007}. NFI is a filtergraph that observed Stokes-$(V/I)$ at a single spectral point of Fe~{\sc i} 6173\ \AA\ line. There was a saturation of Stokes-$(V/I)$ in the magnetic fields exceeding approximately 2000~G. \cite{Chae2007} proposed to use two different linear relationships to evaluate the longitudinal magnetic field from \textit{Hinode}/NFI Stokes-$(V/I)$. Similar analysis was carried out by \cite{Moon2007} who proposed a method to eliminate saturation for magnetic-field measurements performed by \textit{Michelson Doppler Imager} onboard the \textit{Solar and Heliospheric Observatory} \citep[SOHO/MDI:][]{Scherrer1995}. The authors used a second-order polynomial to approximate the strong-field part of SOHO/MDI Stokes-$(V/I)$ versus \textit{Hinode}/SP $B_{\Vert}$ distribution.



\begin{figure}
	\centerline{\includegraphics[width=0.5\textwidth]{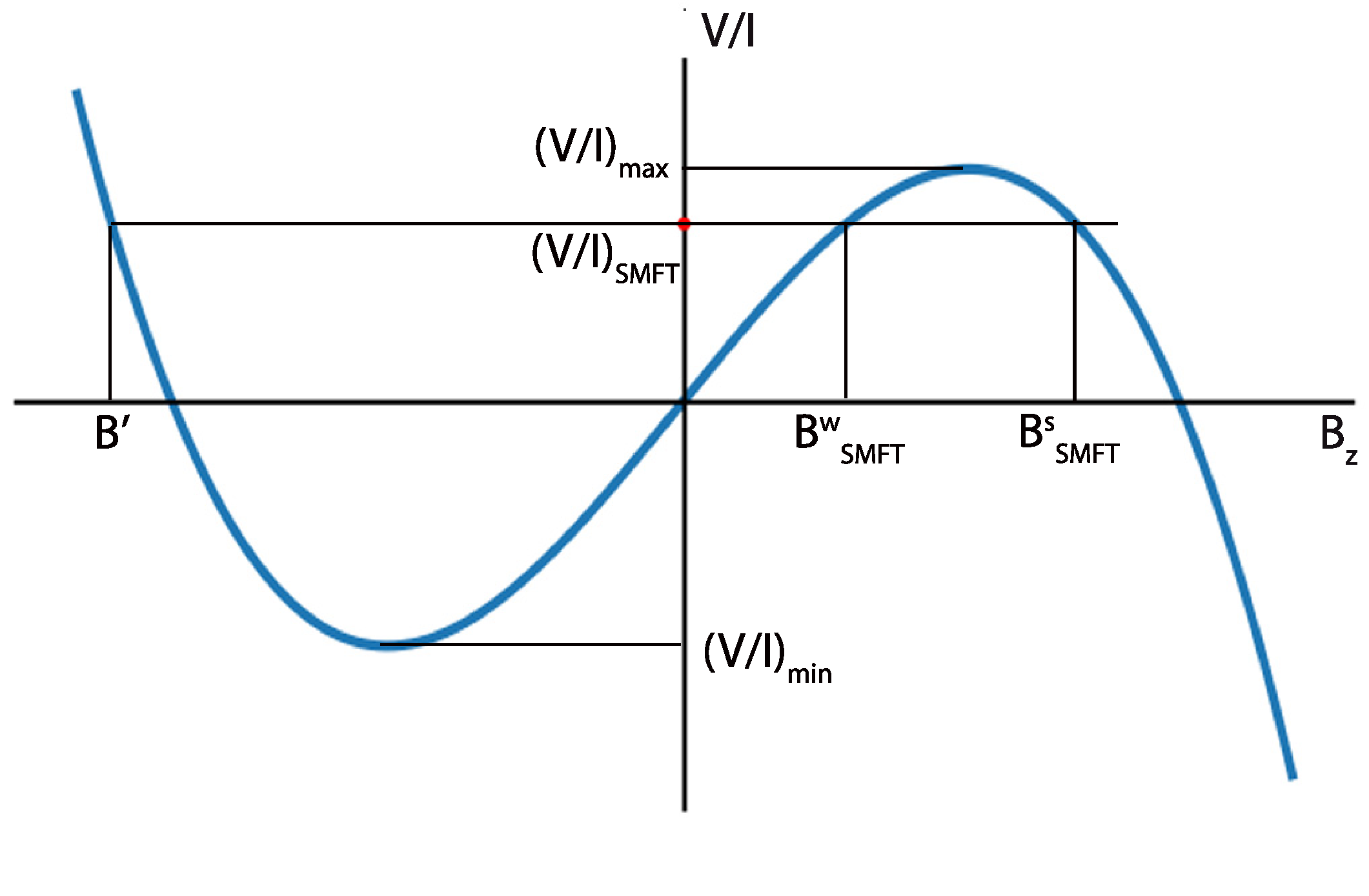}}
	\caption{A schematic illustration of the magnetic-field derivation using the third-order polynomial (Equation \ref{polynom}). The measured Stokes-($V/I$) values $(V/I)_{\rm SMFT}$ correspond to three possible values of the longitudinal magnetic field.}
	\label{BfromVI}
\end{figure}

Figure~\ref{BfromVI} shows schematically the possible way of deriving longitudinal magnetic field from measured Stokes-$(V/I)_{\rm SMFT}$. As it follows from Equation~\ref{polynom}, every value $(V/I)_{\rm SMFT}$ lying between $(V/I)_{\rm min}$ and $(V/I)_{\rm max}$ corresponds to three values of $B_{\rm SMFT}$: $B_{\rm SMFT}^{\rm w}$ (weak-field part of the calibration curve), $B_{\rm SMFT}^{\rm s}$ (strong-field part of the calibration curve), and $B'$. To get these values we solve numerically the equation 
\begin{equation} \label{signal_polynom}
   C_0 + C_1 \times B_{\rm SMFT} + C_2 \times (B_{\rm SMFT})^2 + C_3 \times (B_{\rm SMFT})^3 - (V/I)_{\rm SMFT} = 0,
\end{equation}
The real root of Equation~\ref{signal_polynom} with the sign opposite to that of $(V/I)_{\rm SMFT}$ corresponds to $B'$ and must be rejected. The other two roots are $B_{\rm SMFT}^{\rm w}$ and $B_{\rm SMFT}^{\rm s}$ (the latter has higher absolute value). One of the weak points of the algorithm is the cutoff values labelled by $(V/I)_{\rm max}$ for positive and by $(V/I)_{\rm min}$ for negative $(V/I)_{\rm SMFT}$ in Figure~\ref{BfromVI}.  Apparently, Stokes-$(V/I)_{\rm SMFT}$ values exceeding $(V/I)_{\rm max}$ have to be replaced by $(V/I)_{\rm max}$. Stokes-$(V/I)_{\rm SMFT}$ values that are less than $(V/I)_{\rm min}$ must be replaced by $(V/I)_{\rm min}$ as well. One should keep in mind that this artificial procedure may result in unrealistic morphology of the derived magnetic field maps.

\mdseries

The next step of our method is to figure out at which part of the calibration curve our particular pixel lies, i.e., whether this value of Stokes-$(V/I)_{\rm SMFT}$ in the given pixel corresponds to the weak or the strong magnetic field. To answer this question, we propose to use the information on continuum intensity at the pixel. \cite{Norton2004} showed that there is a direct relationship between continuum intensity and the magnetic field strength. Hence, we can readily use observable Stokes-$I_{\rm SMFT}$ as a proxy of continuum intensity to figure out which part of the calibration curve must be used to derive the value of magnetic field from Stokes-$(V/I)_{\rm SMFT}$: continuum intensity or Stokes-$I_{\rm SMFT}$ in the strong-magnetic-field regions is sufficiently lower than that in the weak-magnetic-field or quiet-Sun areas \citep[cf.][]{Chae2007}.

A straight-forward choice of the threshold value of Stokes-$I_{\rm t}$ for separation of pixels into the two subsets of strong and weak magnetic field might be inappropriate: due to noise in the Stokes-$I_{\rm SMFT}$ maps two adjacent pixels with nearly the same value of the longitudinal magnetic field could be attributed to different parts of the calibration curve (either weak- or strong-field parts). As a result, sudden discontinuities of $B_{\rm SMFT}$ might be observed in the derived map of the longitudinal magnetic field. Another option is to set two thresholds $I_{\rm s}$ and $I_{\rm w}$ ($I_{\rm s} < I_{\rm w}$) such that pixels with $I_{\rm SMFT} < I_{\rm s}$ definitely belong to the strong-field subset and pixels with $I_{\rm SMFT} > I_{\rm w}$ definitely belong to the weak-field subset. In such a case, the magnetic field in pixels with $I_{\rm s} < I_{\rm SMFT} < I_{\rm w} $, i.e. in pixels with some intermediate magnetic fields, has to be derived, for example, by interpolation between the strong and the weak magnetic fields. We tested both approaches and found no advantages of one approach over the other. Both of them may add artificial non-physical structures in the magnetic-field maps. Hence, we use a single threshold $I_{\rm t}$ to distinguish between strong and weak magnetic fields. If the required cadence of the data is not high enough, the manual choice of the threshold value  $I_{\rm t}$ by an observer is acceptable.




The modified algorithm for deriving $B_{\rm SMFT}$ from Stokes-$(V/I)_{\rm SMFT}$ can be summarized as follows:

\begin{enumerate}[i)]
	\item Separate all pixels into two sets of strong and weak magnetic-field pixels by applying threshold $I_{\rm t}$ to Stokes-$I_{\rm SMFT}$ map.
	\item For each pixel of the magnetogram, calculate the roots of the polynomial (Equation \ref{signal_polynom}). If the $(V/I)_{\rm SMFT}$ value in the pixel is greater than $(V/I)_{\rm max}$ (as defined in Figure~\ref{BfromVI}) set $(V/I)_{\rm SMFT} = (V/I)_{\rm max}$. Similarly, pixels with $(V/I)_{\rm SMFT}$ less than $(V/I)_{\rm min}$ has to be replaced by $(V/I)_{\rm min}$.
	\item Set the magnetic-field value in the pixel to a corresponding root of the polynomial: weak-field pixels (as defined in step i) correspond to the root with lower absolute value.
	\item Smooth the magnetogram to eliminate possible discontinuities near pixels with Stokes-$I_{\rm SMFT} = I_{\rm t}$.
\end{enumerate}

\begin{figure}
	\centerline{\includegraphics[width=1\textwidth]{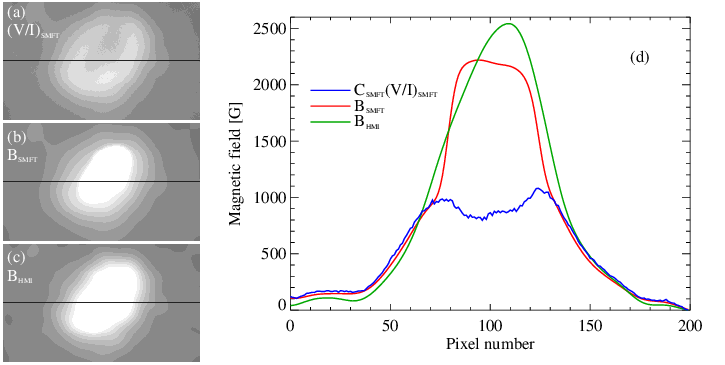}}
	\caption{
	    Maps of Stokes-$(V/I)_{\rm SMFT}$ (a), magnetic field $B_{\rm SMFT}$ derived by the proposed algorithm (b), and smoothed SDO/HMI magnetic field $B_{\rm HMI}$ (c) of NOAA Active Region 12670 observed on 17 August 2017. The field-of-view is 50$^{\prime\prime}\times$35$^{\prime\prime}$. The maps are scaled from -1500\ G (black) to 1500\ G (white). Panel d demonstrates slices of the maps shown in Panels \,a--\,c. The positions of the slices are denoted by horizontal black lines in Panels \,a--\,c.
	}
	\label{fig3}
\end{figure}

Figure~\ref{fig3} demonstrates Stokes-$(V/I)_{\rm SMFT}$ (Panel a), $B_{\rm SMFT}$ derived by the proposed algorithm (Panel b), and $B_{\rm HMI}$ (Panel c) of unipolar NOAA Active Region 12670. A clear saturation of $(V/I)_{\rm SMFT}$ in the active region's umbra is visible. The slices of the maps are shown in Figure~\ref{fig3}d. The saturation in the umbra was eliminated in the final map of $B_{\rm SMFT}$. The maximum $B_{\rm SMFT}$ value inside umbra is about 2200~G which is comparable to that derived by SDO/HMI. For comparison, $B_{\rm SMFT}$ derived from $(V/I)_{\rm SMFT}$ by the linear relationship (Equation \ref{weak-field}) ($C_{\Vert} = C_{\rm SMFT}$) would yield approximately 2.7 times lower values of the magnetic field inside the umbra.

To verify the consistency of our methodology, we derived longitudinal magnetic field $B_{\rm HMI}^{\rm wf}$ from SDO/HMI Stokes-$(V/I)_{\rm HMI}$ measurements acquired at a single wavelength. To do so, we performed the cross-calibration between $(V/I)_{\rm HMI}$ and $B_{\rm HMI}$ to obtain coefficients in Equation \ref{polynom} suitable for the SDO/HMI instrument. Then the $B_{\rm HMI}^{\rm wf}$ map was evaluated by the algorithm described above. Indeed, in contrast to the ground-based HSOS/SMFT, cross-calibration of such reduced SDO/HMI Stokes-$(V/I)_{\rm HMI}$ at a single spectral line point and SDO/HMI-$B_{\rm HMI}$ obtained by VFISV code is free from all errors caused by different seeing, spatial resolution, etc. 

\begin{figure}   
	\centerline{
		\includegraphics{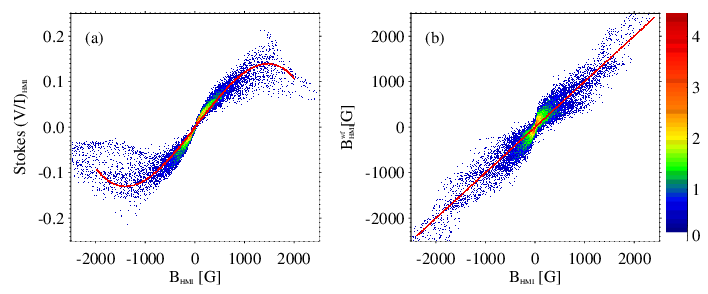}
	}	
	\caption{
	    The logarithmic density of $(V/I)_{\rm HMI}$ versus $B_{\rm HMI}$ and of $B^{\rm wf}_{\rm HMI}$ versus $B_{\rm HMI}$ distributions of NOAA Active Region 12674 observed on 5 September 2017 at 05:48 TAI. $B^{\rm wf}_{\rm HMI}$ is the longitudinal magnetic field derived by the proposed algorithm using Stokes-$(V/I)_{\rm HMI}$ measurements at a single wavelength point. Red curve in Panel a is the best third-order polynomial fit of the distribution. Red curve in Panel b shows $y=x$ relationship.
	}

	\label{fig4}
	
\end{figure}

SDO/HMI Stokes-$(V/I)_{\rm HMI}$ maps were derived from the observations in the fourth filter position. The data on NOAA Active Region 12674 acquired on 5 September 2017 at 05:48 TAI were used. SDO/HMI Stokes-$(V/I)_{\rm HMI}$ versus SDO/HMI-$B_{\rm HMI}$ is shown in Figure~\ref{fig4}a. The same saturation effect in strong magnetic fields as in Figure~\ref{fig1} can be seen. The scatter plot of the derived SDO/HMI longitudinal magnetic field $B_{\rm HMI}^{\rm wf}$ versus SDO/HMI-$B_{\rm HMI}$ is shown in Figure~\ref{fig4}b. One can see a good consistency between the data series. The linear correlation coefficient between $B_{\rm HMI}^{\rm wf}$ and $B_{\rm HMI}$ is 0.96.

\begin{figure}
	\centerline{\includegraphics[width=1\textwidth]{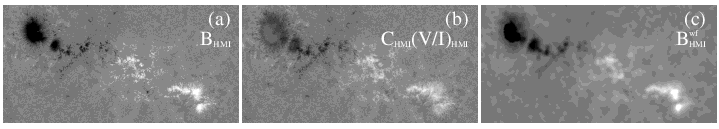}}
	\caption{
		Maps of the longitudinal magnetic filed $B_{\rm HMI}$ (a), Stokes $C_{\rm HMI} (V/I)_{\rm HMI}$ (b), and $B_{\rm HMI}^{\rm wf}$ (c) of NOAA Active Region 12674 acquired by SDO/HMI on 5 September 2017 at 05:48 TAI. $(B)^{\rm wf}_{\rm HMI}$ is the longitudinal magnetic field derived by the proposed algorithm using Stokes-$(V/I)_{\rm HMI}$ measurements at a single wavelength point. The field-of-view is 350$^{\prime\prime}\times$175$^{\prime\prime}$. The magnetic-field values over the magnetograms are scaled from -2000\ G (black) to 2000\ G (white). The constant $C_{\rm HMI}$ is the linear calibration coefficient between $B_{\rm HMI}$ and Stokes-$(V/I)_{\rm HMI}$ (cf. Figure~\ref{fig3}).
	}
	\label{fig5}
\end{figure}

Figure~\ref{fig5} shows $B_{\rm HMI}$, $(V/I)_{\rm HMI}$, and $B_{\rm HMI}^{\rm wf}$ maps of NOAA active region 12674 in Panels a, b, and c, respectively. One can see that the method compensates the saturation effect inside the stronger leading and the weaker following sunspots in the active region.

\section{Results}

\begin{figure}
	\centerline{\includegraphics[width = 1.\textwidth]{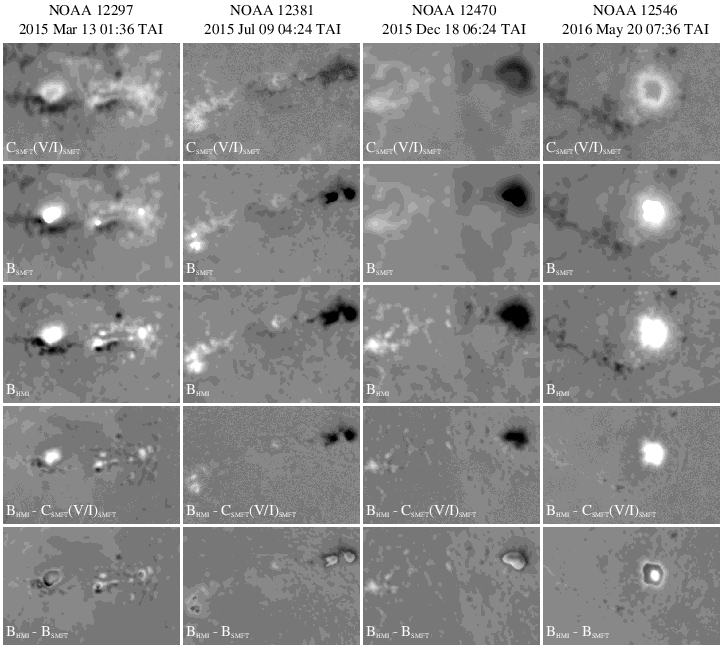}}
	\caption{
		Maps (from top to bottom) of Stokes$- C_{\rm SMFT} (V/I)_{\rm SMFT}$, longitudinal magnetic field $B_{\rm SMFT}$ derived using the proposed algorithm, smoothed SDO/HMI longitudinal magnetic field $B_{\rm HMI}$, the difference $B_{\rm HMI} - C_{\rm SMFT} (V/I)_{\rm SMFT}$, and the difference $B_{\rm HMI} - B_{\rm SMFT}$ for four NOAA Active Regions 12297, 12381, 12470, and 12546 (from left to right). The field-of-view of the maps is 230$^{\prime\prime}\times$155$^{\prime\prime}$. The magnetic field values are scaled from -1500\ G (black) to 1500\ G (white). 
	}
	\label{fig6}
\end{figure}

To illustrate the performance of the method, we derived longitudinal magnetic field $B_{\rm SMFT}$ of four active regions (Figure~\ref{fig6}, second row from the top). Stokes-$(V/I)_{\rm SMFT}$ acquired by HSOS/SMFT and co-temporary SDO/HMI magnetograms of the same active regions are shown in the first and the third rows (from the top) of Figure~\ref{fig6} for comparison. To derive $B_{\rm SMFT}$ the intensity threshold was set to $I_{\rm t} = 0.5 I_{\rm c}$, where $I_{\rm c}$ is Stokes-$I_{\rm SMFT}$ of the quiet-Sun intensity.

One can see that saturation in the Stokes-$(V/I)_{\rm SMFT}$ is observed inside the strong magnetic field concentrations (see the maps in the top row of Figure~\ref{fig6}). At the same time, both the HSOS/SMFT magnetograms derived using the proposed algorithm, and the SDO/HMI magnetograms are free of this effect. The algorithm allows one to reconstruct the morphological structure of an active region. For example, the leading negative polarity of NOAA Active Region 12470 in the $(V/I)_{\rm SMFT}$ map (top row) represents a horseshoe-shaped structure. In the $B_{\rm SMFT}$ map this magnetic feature is a well-defined strong sunspot.

The difference maps between $B_{\rm HMI}$ and $C_{\rm SMFT} (V/I)_{\rm SMFT}$ are shown in the fourth (from the top) row of Figure~\ref{fig6}. A significant difference is observed in the sunspot umbrae implying considerable underestimation of the magnetic-field strength. On the other hand, in the sunspot umbrae the difference maps between $B_{\rm HMI}$ and $B_{\rm SMFT}$ (the bottom row in Figure~\ref{fig6}) in most cases show the sign opposite to that of $B_{\rm HMI} - C_{\rm SMFT} (V/I)_{\rm SMFT}$. Consequently, the proposed algorithm often overestimates the magnetic-field magnitude inside areas with strong magnetic fields. Besides that, the artifacts of the algorithm (\textit{e.g.} concentric rings around sunspot penumbra) can be revealed in the difference $B_{\rm HMI} - B_{\rm SMFT}$ maps. These artifacts could be probably diminished by varying the threshold $I_{\rm t}$ for each active region individually.

The standard deviation of the difference maps shown in the bottom rows of Figure~\ref{fig6} is listed in Table~\ref{tab_stddev}. One can see that the standard deviation of $B_{\rm HMI} - C_{\rm SMFT} (V/I)_{\rm SMFT}$ maps is approximately two times higher than that of $B_{\rm HMI} - B_{\rm SMFT}$ maps. Consequently, in general the proposed algorithm yields better estimation of the longitudinal magnetic field as compared to traditional weak-field approximation.

\begin{table}
\caption{ The standard deviation [G] of the difference maps shown in the bottom rows of Figure~\ref{fig6}. }
\label{tab_stddev}
\begin{tabular}{ccclc}     
  \hline                   
& \multicolumn{4}{c}{NOAA AR} \\
& 12297 & 12381 & 12470 & 12546 \\
  \hline
$B_{\rm HMI} - C_{\rm SMFT} (V/I)_{\rm SMFT}$ & 138 & 125  & 135 & 255\\
$B_{\rm HMI} - B_{\rm SMFT}$                  & 92  & 71   & 84  & 100 \\
  \hline
\end{tabular}
\end{table}

\begin{figure}
	\centerline{\includegraphics[width = 1\textwidth]{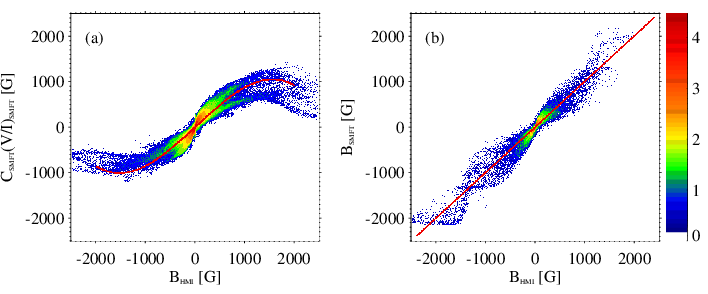}}
	\caption{
	The logarithmic density of $C_{\rm SMFT}(V/I)_{\rm SMFT}$ versus $B_{\rm HMI}$ and of $B_{\rm SMFT}$ versus $B_{\rm HMI}$ distributions for the four active regions shown in Figure~\ref{fig6}. Red curve in Panel a is the best third-order polynomial fit of the distribution. Red curve in Panel b shows $y=x$ relationship. A quasi-linear relationship in Panel b holds for magnetic fields as high as 2000\ G.
	}
	\label{fig7}
\end{figure}

The distributions $C_{\rm SMFT} (V/I)_{\rm SMFT}$ versus $B_{\rm HMI}$ and $B_{\rm SMFT}$ versus $B_{\rm HMI}$ for four selected active regions (Figure~\ref{fig6}) are shown in Figure~\ref{fig7}. In contrast to Figures~\ref{fig1} and \ref{fig7}a, a quasi-linear relationship holds for magnetic fields as high as 2000\ G in the right panel of Figure~\ref{fig7}.

The efficiency of the algorithm can also be evaluated by calculating the correlation coefficient between $B_{\rm HMI}$ and the magnetic field maps derived from HSOS/SMFT observations. However, the distributions in Figure~\ref{fig7} suggest that the data points exhibiting weak-field values are the most numerous. These points may affect the correlation coefficient significantly. To get more reliable value of Pearson's $R$ we split the points in both panels of Figure~\ref{fig7} into ten subsets according to their $B_{\rm HMI}$ values. In each subset, 2000 points were randomly selected. The linear correlation coefficients were calculated for all the selected points that formed equally distributed $C_{\rm SMFT} (V/I)_{\rm SMFT}$ versus $B_{\rm HMI}$ and  $B_{\rm SMFT}$ versus $B_{\rm HMI}$ relationships. This procedure yields Pearson's $R = 0.66$ for $C_{\rm SMFT} (V/I)_{\rm SMFT}$ versus $B_{\rm HMI}$ distribution (Figure~\ref{fig7}a) and Pearson's $R = 0.99$ for $B_{\rm SMFT}$ versus $B_{\rm HMI}$ (Figure~\ref{fig7}b) distribution.


\begin{figure}
	\centerline{\includegraphics{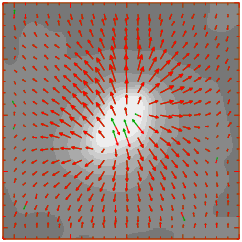}}
	\caption{
		Longitudinal magnetic field $B_{\rm SMFT}$ in NOAA Active Region 12670 acquired by HSOS/SMFT on 17 August 2017 at about 05:18 UT. Red and green arrows point out the direction of the transverse magnetic field for different data on longitudinal magnetic field used for the 180-degree disambiguation (see text). The field-of-view is $45^{\prime\prime}\times45^{\prime\prime}$. The magnetogram is scaled from -2000 G (black) to 2000 G (white).
	}
	\label{fig8}
\end{figure}

Finally, Figure~\ref{fig8} shows transverse magnetic field in NOAA Active Region 12670 that was inferred from the observed HSOS/SMFT Stokes-$Q$ and -$U$. To perform the 180-degree disambiguiation, the transverse magnetic-field vector in each pixel of the magnetogram was co-aligned with the direction of the transversal potential magnetic field. The potential field was calculated from the saturated longitudinal magnetic field (Figure~\ref{fig3}a) for green arrows and from de-saturated longitudinal magnetic field (Figure~\ref{fig3}b) for red arrows. As expected, the saturation prevents transversal magnetic field from being oriented in an incorrect direction inside umbral regions.

\section{Summary}

We used SDO/HMI longitudinal magnetic-field data on several active regions to perform non-linear cross-calibration of HSOS/SMFT Stokes-$(V/I)_{\rm SMFT}$. The essential element of the method is using non-linear relationships between Stokes-$(V/I)$ and the longitudinal magnetic field. Due to certain procedures (for example, setting the limit for $(V/I)_{\rm SMFT} = (V/I)_{\rm max}$ if $(V/I)_{\rm SMFT}$ exceeds $(V/I)_{\rm max}$) the method is not suitable for precise reconstruction of the spatial structure of longitudinal magnetic field from Stokes-$(V/I)_{\rm SMFT}$. However, the main advantage of the method is elimination of saturation inside strong magnetic-field areas such as sunspot umbrae. This feature of the method is important for resolving of 180-degree disambiguation for the transverse magnetic field and subsequent calculation of the electric-current density. Thus, the method might be applied to the archive of HSOS/SMFT vector magnetic-field observations in order to improve the evaluation of the electric currents and current helicity in solar active regions since 1987, which can be implemented in forthcoming studies.

\begin{acks}[Acknowledgments] We are sincerely grateful to the anonymous reviewer whose comments helped us reconsider the method and to significantly improve the article. The  work  is  supported  by  the  joint  cost-shared  RFBR  of  Russia  and  NNSF  of  China  grant, Russian  Fund  for  Basic  Research  numbers 17-52-53203 and 19-52-53045,  also    grants  11427901,  10921303, 11673033, U1731113,11611530679, 11573037, 11703042, 11911530089, 12073040, and 12073041 of the National Natural Science Foundation of China. Theoretical calculations in Section 2 were performed with the support by the Russian Science Foundation, Project 18-12-00131. A.A. Plotnikov, A.S. Kutsenko and K.M. Kuzanyan would like to thank Huairou Solar Observing station of NAOC for their hospitality during their visits to China.\end{acks}

{\footnotesize
\paragraph*{Dislosure of Potential Conflicts of Interest}
The authors declare that they have no coflicts of interest.}


\begin{thebibliography}{}


\bibitem[Ai, Li, and Zhang(1982)]
{Ai1982}
Ai, G.-X., Li, W., Zhang, H.-Q.: 1982, {\it Chinese Astron. and Astrophys.} {\bf 6}, 129.
\href{https://doi.org/10.1016/0275-1062(82)90044-3}{DOI}
\href{https://ui.adsabs.harvard.edu/#abs/1982ChA&A...6..129A/abstract}{ADS}

\bibitem[\protect\citeauthoryear{{Ai}}{1987}]
{Ai1987}
Ai,~G.-X.: 
1987, \textit{Publ. Beijing Astron. Obs.} \textbf{9}, 27.
\href{https://ui.adsabs.harvard.edu/#abs/1987PBeiO...9...27A/abstract}{ADS}


\bibitem[Bai, Deng, and Su(2013)]
{Bai2013}
Bai, X.Y., Deng, Y.Y., and Su, J.T.: 2013, \solphys {\bf 282}, 405.
\href{https://doi.org/10.1007/s11207-012-0197-0}{DOI}
\href{https://ui.adsabs.harvard.edu/#abs/2013SoPh..282..405B/abstract}{ADS}


\bibitem[\protect\citeauthoryear{Bai \emph{et al.}}{2014}]
{Bai2014}
Bai, X.Y., Deng, Y.Y., Teng, F., Su, J.T., Mao, X.J., Wang, G.P.: 2014, {\mnras} {\bf 445}, 49.
\href{https://doi.org/10.1093/mnras/stu1711}{DOI}
\href{https://ui.adsabs.harvard.edu/#abs/2014MNRAS.445...49B/abstract}{ADS}

\bibitem[\protect\citeauthoryear{Bellot Rubio}{2003}]
{BellotRubio2003}
Bellot Rubio, L. R. 2003, Inversion of Stokes profiles with SIR (Freiburg: Kiepenheuer Institut für Sonnenphysik)

\bibitem[Bobra \emph{et al.}(2014)]
{Bobra2014}
Bobra, M.G., Sun, X., Hoeksema, J.T., Turmon, M., Liu, Y., Hayashi, K., and, ...: 2014, \solphys {\bf 289}, 3549. 
\href{https://doi.org/10.1007/s11207-014-0529-3}{DOI}
\href{https://ui.adsabs.harvard.edu/#abs/2014SoPh..289.3549B/abstract}{ADS}


\bibitem[Borrero \emph{et al.}(2011)]
{Borrero2011}
Borrero, J.M., Tomczyk, S., Kubo, M., Socas-Navarro, H., Schou, J., Couvidat, S., and, ...: 2011, \solphys {\bf 273}, 267. \href{https://doi.org/10.1007/s11207-010-9515-6}{DOI}
\href{https://ui.adsabs.harvard.edu/#abs/2011SoPh..273..267B/abstract}{ADS}


\bibitem[\protect\citeauthoryear{Canfield \emph{et al.}}{1993}]
{Canfield1993}
Canfield, R.C., de La Beaujardiere, J.-F., Fan, Y., Leka, K.D., McClymont, A.N., Metcalf, T.R., Mickey, D.L., Wuelser, J.-P., Lites, B.W.: 1993, {\apj} {\bf 411}, 362.
\href{https://doi.org/10.1086/172836}{DOI}
\href{https://ui.adsabs.harvard.edu/#abs/1993ApJ...411..362C/abstract}{ADS}


\bibitem[\protect\citeauthoryear{Chae \emph{et al.}}{2007}]
{Chae2007}
Chae, J., Moon, Y.-J., Park, Y.-D., Ichimoto, K., Sakurai, T., Suematsu, Y., Tsuneta, S., Katsukawa, Y., Shimizu, T., Shine, R.A., Tarbell, T.D., Title, A.M., Lites, B.: 2007, {\pasj} {\bf 59}, S619.
\href{https://doi.org/10.1093/pasj/59.sp3.S619}{DOI}
\href{https://ui.adsabs.harvard.edu/#abs/2007PASJ...59S.619C/abstract}{ADS}




\bibitem[del Toro Iniesta(2003)]
{delToroIniesta2003}
del Toro Iniesta, J.C.: 2003, {\it Introduction to Spectropolarimetry, by Jose Carlos del Toro Iniesta, pp. 244. ISBN 0521818273. Cambridge, UK: Cambridge University Press, April 2003.}, 244.


\bibitem[del Toro Iniesta and Ruiz Cobo(2016)]
{delToroIniesta2016}
del Toro Iniesta, J.C. and Ruiz Cobo, B.: 2016, {\it Liv. Rev. in Solar Phys.} {\bf 13}, 4. \href{https://doi.org/10.1007/s41116-016-0005-2}{DOI}
\href{https://ui.adsabs.harvard.edu/#abs/2016LRSP...13....4D/abstract}{ADS}


\bibitem[Fontenla, Avrett, and Loeser(1993)]
{Fontenla1993}
Fontenla, J.M., Avrett, E.H., and Loeser, R.: 1993, \apj {\bf 406}, 319. 
\href{https://doi.org/10.1086/172443}{DOI}
\href{https://ui.adsabs.harvard.edu/#abs/1993ApJ...406..319F/abstract}{ADS}


\bibitem[\protect\citeauthoryear{Guo \emph{et al.}}{2020}]
{Guo2020}
Guo, J., Bai, X., Deng, Y., Liu, H., Lin, J., Su, J., Yang, X., and Ji, K.: 2020, {\solphys} {\bf 295}, 5. \href{https://doi.org/10.1007/s11207-019-1573-9}{DOI}
\href{https://ui.adsabs.harvard.edu/#abs/2020SoPh..295....5G/abstract}{ADS}


\bibitem[Hoeksema \emph{et al.}(2014)]
{Hoeksema2014}
Hoeksema, J.T., Liu, Y., Hayashi, K., Sun, X., Schou, J., Couvidat, S., and, ...: 2014, {\it Solar Physics} {\bf 289}, 3483.
\href{https://doi.org/10.1007/s11207-014-0516-8}{DOI}
\href{https://ui.adsabs.harvard.edu/#abs/2014SoPh..289.3483H/abstract}{ADS}



\bibitem[Jefferies, Lites, and Skumanich(1989)]
{Jefferies1989}
Jefferies, J., Lites, B.W., and Skumanich, A.: 1989, \apj {\bf 343}, 920.
\href{https://doi.org/10.1086/167762}{DOI}
\href{https://ui.adsabs.harvard.edu/#abs/1989ApJ...343..920J/abstract}{ADS}


\bibitem[\protect\citeauthoryear{Kosugi \emph{et al.}}{2007}]
{Kosugi2007}
Kosugi, T., Matsuzaki, K., Sakao, T. \etal: 2007, {\solphys}, {\bf 243}, 3.
\href{https://doi.org/10.1007/s11207-007-9014-6}{DOI}
\href{https://ui.adsabs.harvard.edu/#abs/2007SoPh..243....3K/abstract}{ADS}


\bibitem[Landi Degl'Innocenti and Landi Degl'Innocenti(1973)]
{LandiDeglInnocenti1973}
Landi Degl'Innocenti, E. and Landi Degl'Innocenti, M.: 1973, \solphys {\bf 31}, 299. 
\href{https://doi.org/10.1007/BF00152807}{DOI}
\href{https://ui.adsabs.harvard.edu/#abs/1973SoPh...31..299L/abstract}{ADS}


\bibitem[Landi Degl'Innocenti and Landolfi(2004)]
{LandiDeglInnocenti2004}
Landi Degl'Innocenti, E. and Landolfi, M.: 2004, {\it Polarization in Spectral Lines. By Egidio Landi Degl'innocenti and Marco Landolfi, University of Fierenze, Firenze, Italy; Arcetri Observatory, Firenze, Italy. ASTROPHYSICS AND SPACE LIBRARY Volume 307 Kluwer Academic Publishers, Dordrecht}.
\href{https://doi.org/10.1007/978-1-4020-2415-3}{DOI}
\href{https://ui.adsabs.harvard.edu/#abs/2004ASSL..307.....L/abstract}{ADS}


\bibitem[Maltby \emph{et al.}(1986)]
{Maltby1986}
Maltby, P., Avrett, E.H., Carlsson, M., Kjeldseth-Moe, O., Kurucz, R.L., and Loeser, R.: 1986, \apj {\bf 306}, 284. \href{https://doi.org/10.1086/164342}{DOI}
\href{https://ui.adsabs.harvard.edu/#abs/1986ApJ...306..284M/abstract}{ADS}



\bibitem[\protect\citeauthoryear{Metcalf}{1994}]
{Metcalf1994}
Metcalf, T.R.: 1994, {\solphys} {\bf 155}, 235.
\href{https://doi.org/10.1007/BF00680593}{DOI}
\href{https://ui.adsabs.harvard.edu/#abs/1994SoPh..155..235M/abstract}{ADS}


\bibitem[\protect\citeauthoryear{Moon \emph{et al.}}{2007}]
{Moon2007}
Moon, Y.-J., Kim, Y.-H., Park, Y.-D., Ichimoto, K., Sakurai, T., Chae, J., Cho, K.S., Bong, S., Suematsu, Y., Tsuneta, S., Katsukawa, Y., Shimojo, M., Shimizu, T.: 2007, {\pasj} {\bf 59}, S625.
\href{https://doi.org/10.1093/pasj/59.sp3.S625}{DOI}
\href{https://ui.adsabs.harvard.edu/#abs/2007PASJ...59S.625M/abstract}{ADS}


\bibitem[Norton and Gilman(2004)]
{Norton2004}
Norton, A.A. and Gilman, P.A.: 2004, \apj {\bf 603}, 348.
\href{https://doi.org/10.1086/381362}{DOI}
\href{https://ui.adsabs.harvard.edu/#abs/2004ApJ...603..348N/abstract}{ADS}



\bibitem[Rachkovsky(1962)]
{Rachkovsky1962}
Rachkovsky, D.N.: 1962, {\it Izv. Krym. Astrof. Observ.} {\bf 28}, 259.
\href{https://ui.adsabs.harvard.edu/#abs/1962IzKry..28..259R/abstract}{ADS}


\bibitem[\protect\citeauthoryear{{Rees} and {Semel}}{1979}]
{Rees1979}
Rees, D.E., Semel, M.D.: 1979, {\aap} {\bf 74}, 1.
\href{https://ui.adsabs.harvard.edu/#abs/1979A&A....74....1R/abstract}{ADS}

\bibitem[\protect\citeauthoryear{{de la Cruz Rodríguez} and {van Noort}}{2017}]
{Rodriguez}
J. de la Cruz Rodríguez and M. van Noort: 2017, {\ssr} {\bf 210}, 109–143.
\href{https://doi.org/10.1007/s11214-016-0294-8}{DOI}


\bibitem[Ruiz Cobo and del Toro Iniesta(1992)]
{RuizCobo1992}
Ruiz Cobo, B. and del Toro Iniesta, J.C.: 1992, \apj {\bf 398}, 375.
\href{https://doi.org/10.1086/171862}{DOI}
\href{https://ui.adsabs.harvard.edu/#abs/1992ApJ...398..375R/abstract}{ADS}


\bibitem[\protect\citeauthoryear{Sakurai \emph{et al.}}{1995}]
{Sakurai1995}
Sakurai, T., Ichimoto, K., Nishino, Y., Shinoda, K., Noguchi, M., Hiei, E., Li, T., He, F., Mao, W., Lu, H., Ai, G., Zhao, Z., Kawakami, S.: 1995, {\pasj} {\bf 47}, 81.
\href{https://ui.adsabs.harvard.edu/#abs/1995PASJ...47...81S/abstract}{ADS}


\bibitem[\protect\citeauthoryear{Scherrer \emph{et al.}}{2012}]
{Scherrer2012}
Scherrer, P.H., Schou, J., Bush, R.I., Kosovichev, A.G., Bogart, R.S., Hoeksema, J.T., Liu, Y., Duvall, T.L., Zhao, J., Title, A.M., Schrijver, C.J., Tarbell, T.D., Tomczyk, S.: 2012, {\solphys} {\bf 275}, 207.
\href{https://doi.org/10.1007/s11207-011-9834-2}{DOI}
\href{https://ui.adsabs.harvard.edu/#abs/2012SoPh..275..207S/abstract}{ADS}


\bibitem[\protect\citeauthoryear{Schou \emph{et al.}}{2012}]
{Schou2012}
Schou, J., Scherrer, P.H., Bush, R.I., Wachter, R., Couvidat, S., Rabello-Soares, M.C., Bogart, R.S., Hoeksema, J.T., Liu, Y., Duvall, T.L., Akin, D.J., Allard, B.A., Miles, J.W.: 2012, {\solphys} {\bf 275}, 229.
\href{https://doi.org/10.1007/s11207-011-9842-2}{DOI}
\href{https://ui.adsabs.harvard.edu/#abs/2012SoPh..275..229S/abstract}{ADS}


\bibitem[\protect\citeauthoryear{{Su} and {Zhang}}{2004}]
{Su2004}
Su, J.-T., Zhang, H.-Q.: 2004, {\cjaa} {\bf 4}, 365.  
\href{https://doi.org/10.1088/1009-9271/4/4/365}{DOI}
\href{https://ui.adsabs.harvard.edu/#abs/2004ChJAA...4..365S/abstract}{ADS}


\bibitem[Unno(1956)]
{Unno1956}
Unno, W.: 1956, \pasj {\bf 8}, 108. \href{https://ui.adsabs.harvard.edu/#abs/1956PASJ....8..108U/abstract}{ADS}


\bibitem[\protect\citeauthoryear{Zhang}{2019}]
{Zhang2019}
Zhang, H.: 2019, {\it Science China Physics, Mechanics, and Astronomy} {\bf 62}, 999601.
\href{https://doi.org/10.1007/s11433-018-9368-x}{DOI}
\href{https://ui.adsabs.harvard.edu/#abs/2019SCPMA..6299601Z/abstract}{ADS}


\bibitem[\protect\citeauthoryear{{Zhang} \emph{et al.}}{2010}]
{Zhangetal2010}
Zhang, Hongqi, Sakurai, T., Pevtsov, A., Gao, Yu, Xu, Haiqing, Sokoloff, D. D., Kuzanyan, K.: 2010, {\mnras {\it Lett}} {\bf 402}, L30.
\href{https://doi.org/10.1111/j.1745-3933.2009.00793.x}{DOI}
\href{https://https://ui.adsabs.harvard.edu/abs/2010MNRAS.402L..30Z/abstract}{ADS}

  
\end{thebibliography}


\end{article} 

\end{document}